# The statistical characteristics of static friction

J. Wang, G. F. Wang*, and W. K. Yuan

Department of Engineering Mechanics, SVL, Xi'an Jiaotong University, Xi'an 710049, China

* E-mail: wanggf@mail.xjtu.edu.cn

**Abstract**

Friction is one of the fundamental issues in physics, mechanics and material science with lots of practical applications. However, the understanding of macroscopic friction phenomena from microscopic aspect is still on the way. In this work, molecular dynamic simulations are performed to investigate the static friction between two planar crystal surfaces. The friction force experienced by each atom is tracked and the statistical characteristics of atomic friction force are illuminated. More importantly, the influences of normal load and temperature on the statistical features are generalized. This study provides a new insight on the micro-states of friction.

# 1 Introduction

Friction is of great importance not only in engineering applications but also in theoretical interest. In macroscopic scale, Amontons and Coulomb summarized the law of friction that the friction force is linearly proportional to the normal load by the friction coefficient [1]. This phenomenological law is proved to be general and has been wildly used in practical applications owing to its simplicity [2]. However, the explanation or understanding of friction phenomena from a microscopic aspect is still far from satisfaction.

With the rapid development in experimental apparatus, such as atomic force microscope (AFM) and friction force microscope (FFM), and large-scale computational capacity, now it is possible to explore friction down to nanoscale and even atomic-scale [3,4]. Both elaborate experiments and molecular dynamics simulations revealed that micro- and nano-scale frictions display some new features, which cannot be characterized by the macroscopic friction law [5]. For examples, the friction of a single nano-sized asperity exhibits stick-slip phenomenon [6]. The coefficient of friction depends on the radius of AFM tip [7]. Atomic simulations demonstrated either a linear relation or a sub-linear relation of friction force to load depending on the commensurability of contact surfaces [8]. Luan and Robinson pointed out that atomic-scale roughness has profound effects on friction [9]. Mo et al. [10] found that the friction force depends linearly on the number of atoms interacting across contact surface. Jiang et al. [11] modified the Lennard-Jones potential by adding a Gaussian shaped potential and simulated the friction between graphene,

MoS$_2$, and black phosphorus.

Much effort has also been devoted to bridging the gap between the atomic-scale friction and the macroscopic friction. Luan and Robbins [12] adopted a hybrid atomistic/continuum method to investigate the friction of two-dimensional rough surfaces. Bhushan and Nosonovsky [13] used the strain gradient plasticity to explain the scale effects in friction. Statistical mechanics provides a powerful method to connect the macroscopic quantities of a large system with the microscopic behaviors of its composed atoms. In fact, some initiative statistical concepts have been used in tribology. Greenwood and Williamson described the topography of a rough surface by a statistical distribution of asperity heights [14]. For a Si tip sliding over a grapheme/a-Si substrate system, Li et al. [15] demonstrated the statistical distribution of atomic friction force, but did not provide a detailed mathematical analysis.

In this paper, the static friction between two crystal slabs with large contact surface will be simulated by the embedded atomic method. Then the microscopic statistical characteristic of atomic friction force will be analyzed based on the normal distribution. Furthermore, the influence of normal load and temperature on the probability distribution function of atomic friction force is calculated and generalized.

**2 Simulation Methods**

In this work, molecular dynamics (MD) simulations are performed to investigate the static friction between two planar crystal slabs, as depicted in Fig. 1. Two-dimensional atomic model is considered with two slabs composed of hexagonal-packed copper atoms. The height of each slab is 2.4 nm, and the length of the slabs is about 0.51 μm. Periodic boundary condition is applied in the horizontal direction. The MD simulation package, Large-scale Atomic/Molecular Massively Parallel Simulation (LAMMPS) [16], is used to carry out the atomic simulations.

We utilize the embedded atom method (EAM) [17] to describe the interactions between copper atoms within one slab. For an atomic system containing $N$ atoms, the potential energy $E_b$ of the atom-$i$ is given by

$$E_b = F(\rho_i) + \frac{1}{2} \sum_{i \neq j}^{N} \phi(r_{ij}), \tag{1}$$

where $F(\rho_i)$ is the embedded energy related to the electron density $\rho_i$ at the position of the $i$-th atom, $\phi(r_{ij})$ is the pair potential interaction, and $r_{ij}$ represents the distance between atom pair $i$ and $j$. A potential for copper parameterized by Mishin et al [18] is adopted here, which has been widely used to investigate mechanical properties and deformation mechanisms in various nanostructures.

The atoms in one slab interact with the atoms in another slab through a typical Lennard-Jones potential

$$E_s = 4\varepsilon_s \left[ \left( \frac{\sigma_s}{r_{ij}} \right)^{12} - \left( \frac{\sigma_s}{r_{ij}} \right)^{6} \right], \tag{2}$$

where $\varepsilon_s$ governs the interaction intensity and $\sigma_s$ is the equilibrium spacing. In this work, most of the simulations are carried out with $\varepsilon_s = 0.02$ eV, $\sigma_s = 3.416$ Å and the cut-off radius $r_c = 8$ Å.

The simulations are performed in the framework of microcanonical ensemble (NVT), and the temperature of system is adjusted by the Nosé-Hoover thermostat [19]. After construction of the atomic system, the conjugate gradient algorithm is adopted to relax the whole system in order to reach an equilibrium state. Then the lowest atom layer of the bottom slab is fixed, while the upper slab is moveable. To implement the friction process, two harmonic springs are employed, as shown in Fig. 1. The vertical spring is utilized to apply a normal load $L$. Denote the contact area by $A$, then the average normal pressure $P$ equals $L/A$. The lateral spring is employed to drive the upper slab forward with one end tied at the mass center of the upper slab and another end moving along the horizontal direction. The stiffness constants of the vertical and horizontal springs are 0.01 and 0.02 eV/Å$^2$, respectively. After adding a certain normal force, the atomic system is relaxed again. Then the upper slab is pulled forward by the lateral spring with a constant speed $v$ of 0.02 Å/ps. When the force of the lateral spring goes up beyond the maximum static friction force, the upper slab starts to slide against the bottom one.

In order to study the micro-state of static friction, we will examine the friction force experienced by each atom, named atomic friction force, which is defined as the resultant lateral force exerted on one atom in the upper slab by all atoms in the bottom slab. Summation of the atomic friction force $f$ over all contact atoms yields the ordinary overall friction force $F_{sum}$ between the two slabs.

To eliminate the influence of material selection, all the quantities are scaled to the dimensionless ones by using the parameters related to the single crystal copper. The friction force $f$, normal pressure $P$, temperature $T$, interaction intensity $\varepsilon_s$, the driving force $F_{spr}$ of spring and the overall friction force $F_{sum}$ are normalized as

$$f^* = \frac{f\sigma_b}{\varepsilon_b}, \quad P^* = \frac{P\sigma_b^3}{\varepsilon_b}, \quad T^* = \frac{kT}{\varepsilon_b}, \quad \varepsilon^* = \frac{\varepsilon_s}{\varepsilon_b}, \quad F_{spr}^* = \frac{F_{spr}\sigma_b}{\varepsilon_b}, \quad F_{sum}^* = \frac{F_{sum}\sigma_b}{\varepsilon_b}, \quad (3)$$

where $k$ is the Boltzmann constant, $\varepsilon_b$ is 0.423 eV obtained by fitting the cohesive energy of single crystal copper, $\sigma_b$ is 2.325 Å determined by considering the lattice structure of copper atoms [18]. Thus $\varepsilon_b/\sigma_b$ is about 0.29 nN and $\varepsilon_b/\sigma_b^3$ is about 5.4 GPa.

## 3 Results and discussions

At first, we present the variation of the spring force with the time evolving under a normal pressure $P^* = 0.28$ and temperature $T = 20$ K, as shown in Fig. 2(a). In the stage of static friction, the spring force increases linearly with time. When the spring driving force reaches the maximum static friction force, the upper slab will move and the spring force drops rapidly and almost vanishes shortly after. Different from the stick-slip friction process experienced by one atom or a small cluster of atoms, the friction between two large surfaces as considered in the present paper involves so many atoms that the friction force almost vanishes once the maximum static friction is overcome. In all our simulated cases, such basic features of friction force have been observed.

In order to explore the microscopic statistical characteristic of static friction, we consider four equilibrium states (marked as **A**, **B**, **C**, **D** in Fig. 2(a)) before the maximum friction force has been reached. In each state, we keep the lateral force of spring $F_{spr}^*$ as constant and equilibrate the atom system for an enough long time, so that an equilibrium state of static friction is obtained. Fig. 2(b) records the sum of friction force experienced by all atoms. It seems that the overall friction force $F_{sum}^*$ fluctuates around the applied spring force owing to thermal fluctuation. Based on these equilibrium states, we examine the statistical features of atomic friction force.

Now we consider the atomic friction force experienced by two outmost atom layers in the upper slab. The distribution of atomic friction force at the state **D** is displayed in Fig. 3(a). It seems that, for the friction between two crystal surfaces, the outmost atom layer gives the dominant contribution to the overall friction force, while the contribution of inner atoms can be neglected. Therefore, we concern only the friction force of the outmost atom layer. To reveal the statistical features of atomic friction force, the atom number in different range of friction force is counted as shown in Fig. 3(b). It is interesting to find that the friction force exhibits a normal distribution at atomic scale, with larger population around its central value and much smaller population on the two sides. The probability distribution function of the atomic friction force can be characterized by

$$\rho(f^*) = \frac{1}{\sqrt{2\pi}w^*}\exp\left[-\frac{(f^* - f_c^*)^2}{2(w^*)^2}\right], \qquad (4)$$

where $f_c^*$ and $w^*$ are the mean value and the standard deviation, respectively. For the four considered states, the value of $f_c^*$ and $w^*$ are listed in Table (1). Since $F_{sum}^*$ is equal to $F_{spr}^*$, the mean value $f_c^*$ will depend only on the driving force in the static friction. Table (1) displays that $w^*$ is nearly a constant independent of the driving force.

The distribution of atomic friction force is not a standard normal distribution. To indicate the deviation from a normal distribution, the skewness $S = \mu_3 / (\mu_2)^{3/2}$ and the excess kurtosis $K = \mu_4 / (\mu_2)^2 - 3$ are also given in Table (1), where $\mu_2$, $\mu_3$ and $\mu_4$ are the second, third and fourth central moment. The small positive values of $S$ and $K$ mean that the distribution of atomic friction force is slightly right-skewed and leptokurtic, respectively.

For a specific lateral spring force $P^* = 0.28$, we also check the distribution of atomic friction force at various temperatures $T = 20$ K, 100 K, 200 K and 300 K as shown in Fig. 4. As the temperature rises, both the standard deviation and the kurtosis increase, indicating that the dispersion width of atomic friction force is expanding.

For static friction, the critical state (denoted by the state **E** in Fig. 2(a)), which yields the maximum friction force and thus gives the static frictional coefficient, is always concerned. We will investigate the influence of normal pressure and temperature on the statistical distribution of atomic friction force at such critical state. $f_c^*$ in Eq. (4) is replaced by $f_m^*$ to indicate the critical state. In this way, the mean friction force $f_m^*$ and the standard deviation $w^*$ are two functions of temperature and normal pressure.

At a specific temperature $T = 10$ K, we calculate the probability distribution functions of atomic friction force with increasing normal pressures $P^*$, as shown in Fig. 5(a). It is seen that the mean atomic friction force $f_m^*$ increases with the normal pressure increasing. To get a quantitative analysis, we further determine the mean atomic friction force $f_m^*$ versus the normal pressure, as plotted in Fig. 6(a). It is found that $f_m^*$ is linearly proportional to the normal pressure $P^*$. We also present the corresponding results for two other temperature $T = 50$ K and 150 K. It reveals that the dependence of $f_m^*$ on normal pressure and temperature can be described by such a function

$$f_m^*(T^*, P^*) = g_1(T^*)P^* + g_2(T^*), \tag{5}$$

where $g_1$ is the friction coefficient, defined as the ratio between the friction force $f_m^*$ and the normal pressure $P^*$, $g_2$ is the friction force induced by adhesion. Furthermore, we calculate the variations of $g_1$ and $g_2$ with respect to temperature $T^*$, as shown in Fig. 6(b). By fitting $g_1$ and $g_2$ against temperature, we finally get the mean atomic friction force as

$$f_m^*(T^*, P^*) = (-1.14T^* + 0.146)P^* + (-0.448T^* + 0.0476). \tag{6}$$

This relationship regenerates the macroscopic friction law that the friction force is linear with the normal pressure. Consider the first term in the right hand of Eq. (6), the coefficient of friction for zero temperature is 0.146, which is reasonable for metals, and the coefficient of friction declines with the increasing of temperature.

By the similar way, the standard deviation $w^*$ of atomic friction force can also be calculated with respect to normal pressure and temperature, as shown in Fig. 7(a). The standard deviation $w^*$ is also linear to the normal pressure as

$$w^*(T^*, P^*) = h_1(T^*)P^* + h_2(T^*), \tag{7}$$

And the variations of two coefficient $h_1$ and $h_2$ with respect to temperature are displayed in Fig. 7(b). By fitting the two coefficient $h_1$ and $h_2$, then we obtain the standard deviation $w^*$ as

$$w^*(T^*, P^*) = \sqrt{T^*}(0.66P^* + 0.27). \tag{8}$$

Thus, by substituting Eq. (6) and (8) into Eq. (4), the probability distribution at the state with maximum static friction force can be expressed by

$$\rho(f^*) = \frac{1}{\sqrt{2\pi T^*}(0.66P^* + 0.27)} \exp\left[-\frac{(f^* - (1.2P^* + 0.40)(0.12 - T^*))^2}{2T^*(0.66P^* + 0.27)^2}\right]. \quad (9)$$

From Eq. (9), we can clearly see the micro-states of frictional force experienced by atoms, and the influence of normal pressure and temperature on the statistical characteristics of atomic friction force. This information is helpful to understand friction phenomena at atomic scale and establish the bridge from the micro-states to macroscopic properties related to friction.

## 4 Conclusions

In summary, molecular dynamic simulations are performed to investigate the static friction between two crystal slabs with large contact surface, and its microscopic statistical characteristics are revealed. It is found that the distribution of atomic friction force is approximate to a normal distribution. Through a large amount of simulations, the influence of normal pressure and temperature on the probability distribution function of atomic friction force is characterized. The mean value of atomic friction force rises linearly to the normal pressure, but decreases linearly with respect to temperature. Meanwhile, the standard deviation of atomic friction force is linearly proportional to the normal pressure and the square root of temperature. This work is helpful to understand friction phenomena at atomic scale and establish the bridge from the micro-states to macroscopic friction phenomena.

**Acknowledgments**

Support from the National Natural Science Foundation of China (Grant No. 11525209) is acknowledged.

**References**

[1]. Dowson. D.: History of tribology. Longman Inc. (1998)

[2]. Popov, V.L.: Contact mechanics and friction. Springer Berlin Heidelberg. (2010)

[3]. Braun, O.M., Naumovets, A.G.: Nanotribology: Microscopic mechanisms of friction. Surf. Sci. Rep. **60**(6-7), 79-158 (2006)

[4]. Dong Y., Li Q., Martini A.: Molecular dynamics simulation of atomic friction: A review and guide. J. Vacu. Sci. Tech. A **31**(3), 030801 (2013)

[5]. Carpick, R.W., Salmeron, M.: Scratching the surface: Fundamental investigations of tribology with atomic force microscopy. Chem. Rev. **97**(4), 1163-1194 (1997)

[6]. Li Q., Dong Y., Perez D., Martini A., Carpick R. W.: Speed dependence of atomic stick-slip friction in optimally matched experiments and molecular dynamics simulations. Phys. Rev. Lett. **106**(12), 126101 (2011)

[7]. Yoon, E.S., Singh, R.A., Oh, H.J.: The effect of contact area on nano/micro-scale friction. Wear. **259**(SI), 1424-1431 (2005)

[8]. Wenning, L., Muser, M.H.: Friction laws for elastic nanoscale contacts. Europhys. Lett. **54**(5), 693-699 (2001)

[9]. Luan B.Q., Robbins M.O.: Contact of single asperities with varying adhesion:

Comparing continuum mechanics to atomistic simulations. Phys. Rev. E. **74**(2), 026111 (2006)

[10]. Mo, Y., Turner, K.T., Szlufarska, I.: Friction laws at nanoscale. Nature. **457**(7233), 1116-1119 (2009)

[11]. Jiang, J.W., Park, H.S.: A Gaussian treatment for the friction issue of Lennard-Jones potential in layered materials: Application to friction between graphene, MoS2, and black phosphorus. J. Appl. Phys. **117**(12), 463-300 (2015)

[12]. Luan, B.Q., Robbins, M.O.: Hybrid atomistic/continuum study of contact and friction between rough solids. Tribol. Lett. **36**(1), 1-16 (2009)

[13]. Bhushan B., Nosonovsky M.: Scale effects in friction using strain gradient plasticity and dislocation-assisted sliding. Acta Mater. **51**(14), 4331-4345 (2003)

[14]. Greenwood J.A., Williamson J.B.P.: Contact of nominally flat surfaces. Proc R Soc London A. **295**(1442), 300-319 (1966)

[15]. Li, S.Z.; Li, Q.Y.; Carpick, R.W.: The evolving quality of frictional contact with graphene. Nature. **539**(7630), 541-546 (2016)

[16]. Plimpton, S.: Fast parallel algorithms for short-range molecular dynamics. J. Comput. Phys. **117**(1), 1-19 (1995)

[17]. Daw, M.S., Baskes, M.I.: Embedded-atom method: derivation and application to impurities, surfaces, and other defects in metals. Phys. Rev. B. **29**(12), 6443 (1984)

[18]. Mishin, Y., Mehl, M. J., Papaconstantopoulos, D. A., Voter, A. F., Kress, J. D.: Structural stability and lattice defects in copper: ab initio, tight-binding, and

embedded-atom calculations. Phys. Rev. B. **63**(22), 224106 (2001)

[19]. Nosé, S.: A unified formulation of the constant temperature molecular-dynamics methods. J. Chem. Phys. **81**(1), 511-519 (1984)

**Figure captions**

Fig. 1. The illustration of friction between two crystal slabs.

Fig. 2. The driving force and the overall friction force. (a) The evolving of spring driving force with time. (b) The comparisons between $F_{spr}^*$ and $F_{sum}^*$ at several states.

Fig. 3 The distribution of atomic friction force (a), and its characteristic charts (b).

Fig. 4. The statistical distribution of atomic friction force at various temperatures.

Fig. 5. The distribution of atomic friction force with increasing normal pressures.

Fig. 6 The variation of the mean atomic friction force $f_m^*$ with respect to normal pressure (a), and the dependence of coefficient $g_1$ and $g_2$ on temperature (b).

Fig. 7 The variation of the standard deviation variation $w^*$ with respect to normal pressure (a), and the dependence of coefficient $h_1$ and $h_2$ on temperature (b).

**Table captions**

Table 1. The statistical parameters of atomic friction forces in states **A** to **D**

Table 1

| State | $f_c^*$ | $w^*$ | $S$ | $K$ |
|---|---|---|---|---|
| **A** | 0.034 | 0.029 | 0.10 | 1.15 |
| **B** | 0.050 | 0.028 | 0.26 | 0.41 |
| **C** | 0.065 | 0.027 | 0.46 | 0.40 |
| **D** | 0.076 | 0.026 | 0.46 | 0.50 |

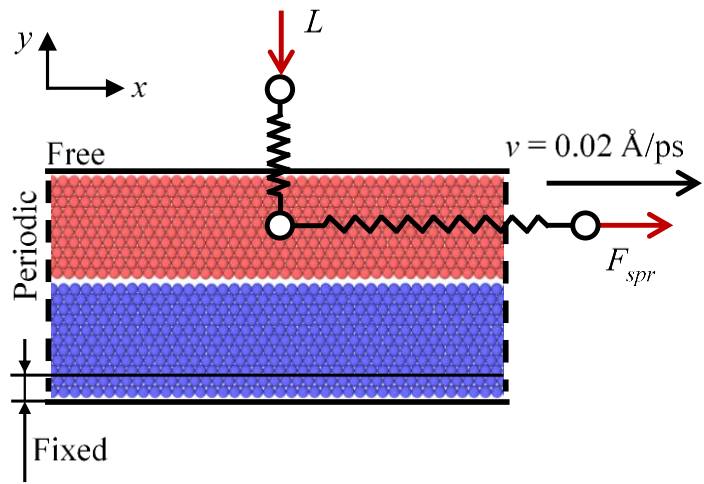

Figure 1

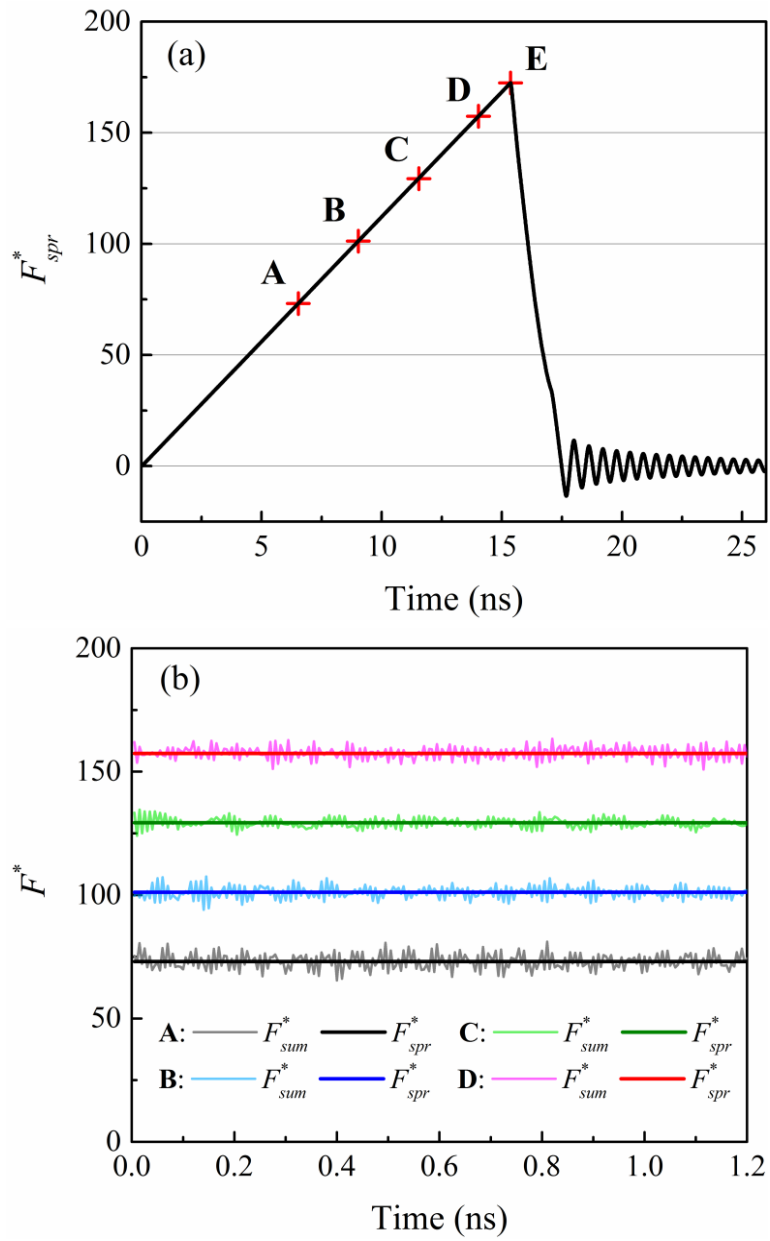

Figure 2

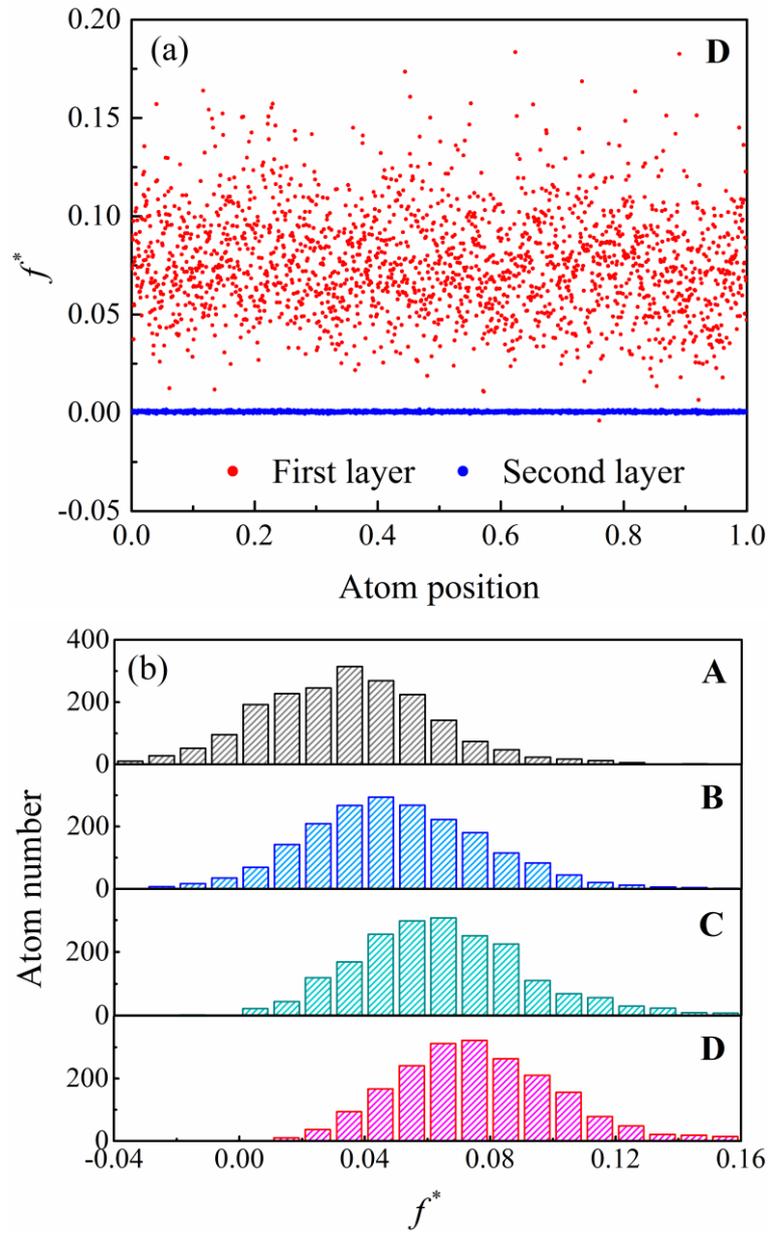

Figure 3

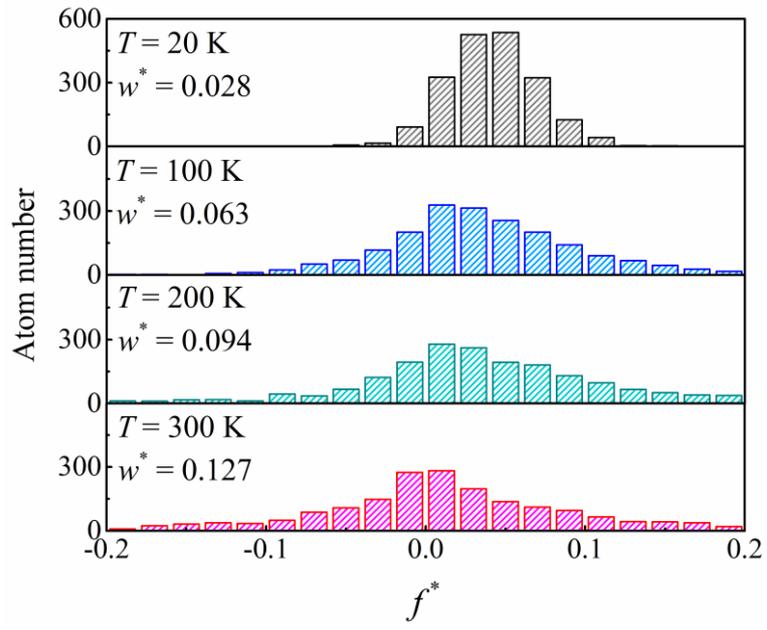

Figure 4

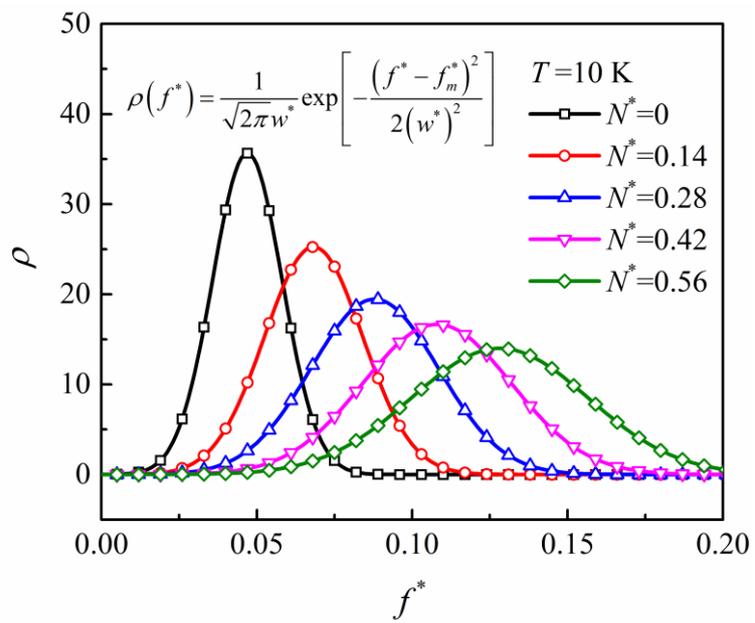

Figure 5

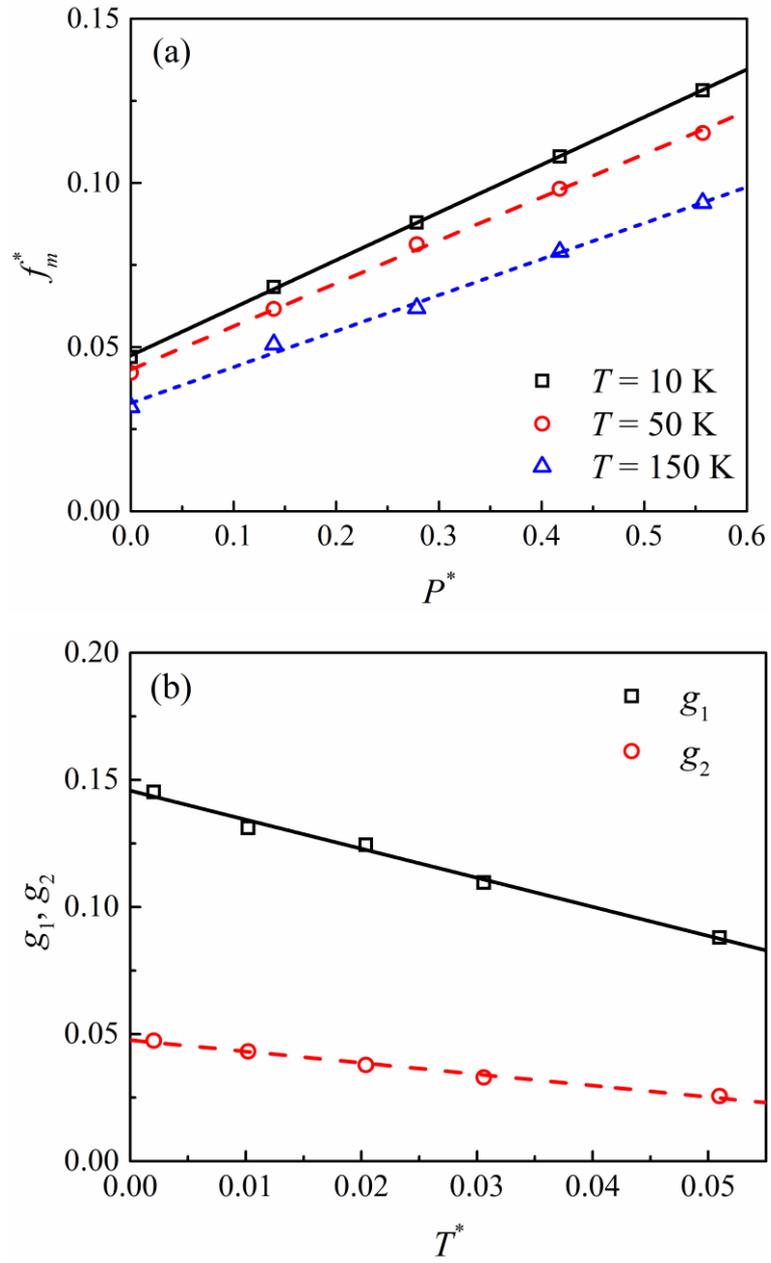

Figure 6

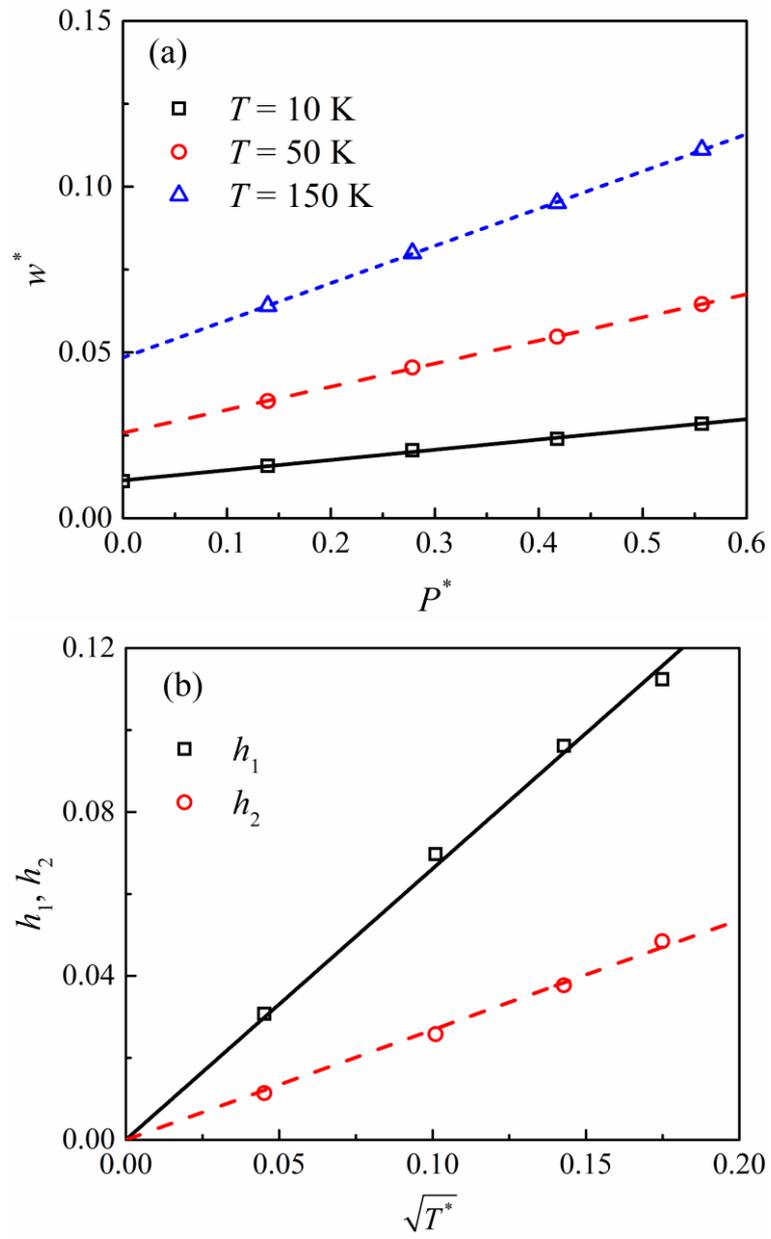

Figure 7